# Web Annotation as a First Class Object

Paolo Ciccarese, Stian Soiland-Reyes and Tim Clark

*Scholars have made handwritten notes and comments in books and manuscripts for centuries. Today's blogs and news sites typically invite users to express their opinions on the published content; URLs allow web resources to be shared with accompanying annotations and comments using third-party services like Twitter or Facebook. These contributions have until recently been constrained within specific services, making them second-class citizens of the Web.*

*Web Annotations are now emerging as fully independent Linked Data in their own right, no longer restricted to plain textual comments in application silos. Annotations can now range from bookmarks and comments, to fine-grained annotations of a selection of, for example, a section of a frame within a video stream. Technologies and standards now exist to create, publish, syndicate, mash-up and consume, finely targeted, semantically rich digital annotations on practically any content, as first-class Web citizens. This development is being driven by the need for collaboration and annotation reuse amongst domain researchers, computer scientists, scientific publishers, and scholarly content databases.*

## Freeing Annotation from Content Silos

Content annotation in one form or another forms a basic part of Web 2.0. **Flickr** lets users tag and categorize others' photos, and to highlight portions of a picture; **Twitter** supports tagging in the form of hashtags, recently also introduced by **Facebook**; **SoundCloud** lets music fans indicate their appreciation at particular points in a song; **Youtube** allows a video to be posted as a response to another video.

Social networking sites like **Facebook** and **Google+** let users share a URL to, say, a news story, enabling a stream of comments within their group of friends, which indicate their "likes". Social media sites like **Reddit** rank lists of news items solely with up/down voting from their users, and the stream of associated discussions provide anything from amusing side-jokes to detailed reviews by experts in the field.

What do these applications have in common? First, that the annotations they support have a shared subject, typically identified by an URI (e.g. to a webpage, image or video); and second, these annotations live within a single system, either at the provider of the resource or at a third-party bookmarking site. Although many of these systems provide REST APIs for third-party access, the APIs differ among sites. Up until very recently there has been no common annotation model independent of what is being annotated. It is therefore difficult for, say, a SoundCloud page to embed comments made within Facebook or Reddit.

While this may not pose a significant problem for ordinary users of social media, it is a big problem in digital scholarship. This problem is driving a significant change in the way annotations can now be implemented and reused across systems and domains.

## Semantics, Not Just Comments

In July 2013, Google announced the availability of 800 million documents annotated by 11 billion Freebase concepts [1]. This is semantic tagging. Text in a document is mapped to a stable URI that identifies a concept, entity, person, place, process, etc, with which it is associated. Information derived from elsewhere – say Wikipedia – is refocused around a common URI in an organized way. In Google's case, this was done using resources in Freebase. But we can also use entries in DBPedia, or other resources, to identify terms. Additionally we can use ontologies: formal vocabularies with classes, properties, and relationships.

Using formal vocabularies specified in the Web Ontology Language (OWL), we can create richer, more computationally tractable structures than we can do using dictionaries or encyclopedias. Formal ontologies relate elements in a domain one to another, using formal predicates, and allow us to perform automated reasoning over these knowledge structures. They may also supplement, identify, or tag dictionary and encyclopedia entries themselves.

If I know that the URI of an ontology term for the gene BACE1 corresponds to some text, the ontology can also tell me what protein is the product of that gene (beta site APP cleaving enzyme), its amino acid sequence, species variants, - and that it is implicated in the pathology of Alzheimer Disease. I can find experimental data annotated with this term, or one of its related terms, as well. And I can assemble mash-ups of related useful information from multiple databases, and overlay them on the text referring to my original gene name. With scientific publication volumes growing exponentially, this kind of machine assistance to reading, comprehension and knowledge accumulation is increasingly necessary.

Semantic tagging or semantic linking is now a reality, and extends beyond Google and Freebase. Perhaps its most important application is in scholarship. For example, the **Europe PubMed Central** (http://europepmc.org/) database of scientific articles now semantically tags all its entries with protein, gene and chemical compound names found in the text. But dealing with these tags raises further important questions.

How do we represent tags, and other annotations? Are they injected in the text, or stored separately? How do we share them across systems? Supposing I annotate a locally-served PDF copy of some text, using comments, semantics tags, and associated video – and I wish to share that annotation with, say, the publisher of the original HTML version of the text. How do I do that? And how can we make this representation live as linked data on the Web – as a first-class citizen? Can I filter and authorize access by author or group? These are a few of the capabilities just becoming available to Web developers using Linked Data Annotation tools and models.

## The Evolution of Web Annotation

### *Origins – Digital Hypermedia to Annotea*

More than a decade ago, seminal lines of research in **Distributed Link Services** (DLS) [2] and conceptual open hypermedia (**COHSE**) [3] explored this area. Their main goal was to enhance collaboration via shared metadata-based Web annotations, bookmarks, and their combinations. Developers of DLS and COHSE drew upon extensive prior research experience in distributed hypermedia systems, predating the Web.

Annotea [4] was a relatively early W3C project to enable collaboration by sharing annotations and metadata, which were attached to a full or partial web page. This was done without modifying the document source, by injecting targeted elements from annotation stores. When a web page was accessed, annotations of that page could be loaded from and published to separate Annotea servers, one of which was hosted by W3C (now decommissioned). Annotea consisted of an RDF schema with an accompanying REST API, for discovery and publication of annotations.

The Annotea data model was lightweight. Annotations were stored separately from the documents they annotated, and consisted of three parts. An *Annotation* associated a Web resource with a *body* - another Web resource - for example, a comment or reply (typically XHTML embedded as an XML Literal). The *Annotation* had provenance properties like *author, created* and *modified*, specializing Dublin Core Elements. Finally a *context* indicated the specific selection within an annotated document, typically using an XPointer. The Annotation class could be subclassed, e.g. *Question, Comment, Example*. Annotea faced however several drawbacks, perhaps most importantly, lack of a driving non-technical user community which required its features to accomplish their goals. But it laid out a basic path that subsequent systems followed and exploited.

### *The Open Annotation Model*

Building on RDF and partially inspired by Annotea, in 2009 two parallel projects were begun independently: the **Open Annotation Collaboration** (OAC) and the **Annotation Ontology** (AO). Both models adopted the fundamental structure of Annotea: an *Annotation* with a *target* (annotated resource) and a *body* (content), which is about the target. They emphasized different functionalities, according to requirements of their user communities – which were strongly coupled to these development projects and drove them.

OAC was designed to support digital humanities, while AO was geared to biomedical use cases, with a core goal of annotating and semantically relating database entries, documents and biomedical concepts from ontologies. Both models were designed to be orthogonal to the domain of interest. OAC and AO took similar-enough approaches to allow merger into a single project, after discussions at a summer workshop convinced both teams that this was possible and desirable.

The **Open Annotation Model** (OA) developed from these prior projects, under the umbrella of the *W3C Open Annotation Community Group* (http://www.w3.org/community/openannotation/). The OA model incorporates features from its predecessors, but with further flexibility and richness, catering for a wider range of annotation needs. It includes use-case and technical contributions from over fifty organizations, and nearly one hundred individual and organization representatives. Open Annotations is currently the fifth largest W3C Community Group by number of participants.

In OA, annotations can be related to the Linked Data cloud either by marking the body as a *semantic tag*; or by using a body that is a dereferencable RDF resource, or, in serializations that support it, an embedded *named graph* (typed as *trig:Graph*). The two last approaches have been adapted by projects like Wf4Ever **Research Objects** (http://www.researchobject.org/) to organize annotation graphs containing detailed relations between resources within the research object, and at the same time to distinguish between annotations coming from different sources. Used in this manner, the annotation model allows a high-level, vocabulary-neutral expression of relationships between resources, with the domain-specific details unrolled within the graph of the body.

## Linked Data Annotation Tools

Several systems already use annotations to produce Linked Data.

**DBpedia Spotlight** (http://spotlight.dbpedia.org/) is a tool for automatically annotating mentions of DBpedia resources in text, providing a solution for linking unstructured information sources to the Linked Open Data cloud through DBpedia.

**Domeo Annotation Toolkit** (http://www.annotationframework.org/) allows users to trigger external text mining services/algorithms and to transform their results into annotations that may be further curated by human agents. Domeo also supports manual, semantically structured annotations using established vocabularies, and can create relational database entries as well as Linked Data.

**Utopia Documents** (http://getutopia.com/) is a PDF reader that when opening a research article can overlay manipulatable visualizations of the data behind the paper, providing data analysis tools and links to external resources. This is made possible through a combination of automatic analysis and manual annotations on elements of the paper, like figures and tables. Numerous Linked Data sources are consulted in the process, including DBPedia and Open PHACTS (http://explorer.openphacts.org/). The tool has been adopted by the Biochemical Journal, where annotations are created by the Journal's editors.

Another valuable example of annotation and Linked Data is **Maphub** (https://maphub.herokuapp.com/), an online application for exploring and annotating digitized, high-resolution historical maps. Maphub allows annotations to be marked with Wikipedia entries for locations and institutions, forming semantic tags to corresponding DBPedia entries.

### Convergence

The Open Annotations model is new. We have never before had a comprehensive, fully-featured model for linked data annotations, with significant community buy-in. As applications and annotation communities fully adopt this model for their work, linked data annotations will become first-class, fully interoperable Web citizens: primary Web content in their own right.

BOXED AREA

***Open Annotation model***

The Open Annotation Model **[5]** is a very rich model, with a simple foundation.

OA specifies instances of an *Annotation,* with predicates *hasTarget,* the annotated resource; and *hasBody*, a resource concerning the target. Further properties indicate the *Annotation's* provenance and motivation. The body and target may be of any type, however OA recommends lightweight typing using DCMI Types to indicate the abstract media type (e.g. *Image*, *Sound, Text*). These can be useful for rendering the annotation intuitively.

The body may be typed as a *Tag*. For Linked Data integration, OA provide the subclass *SemanticTag*, where the body would be specified as a URI to the tagging concept or entity. For ordinary text comments, OA use the Content in RDF specification.

It is possible to specify the Body or Target through indirection with a *SpecificResource*, which indicates the resource URI using *hasSource*. Further constraints may resolve a specific area, segment, etc., applied using a *Selector* (Text selection, SVG area, URI fragments) and *State* (HTTP headers, time) instances. Thus it is possible, for example, to annotate a circular selection of a JPEG image retrieved using content negotiation. The annotation can be *styledBy* a *Style*, e.g. CSS. This, combined with the selector, can be used to specify highlighting of selected text in a particular color, or to play a video in half speed at a certain time segment. Selectors, state and style can be extended for application-specific requirements.

## *About the authors*

Paolo Ciccarese (http://orcid.org/0000-0002-5156-2703) is an Instructor of Neurology at Harvard Medical School, Assistant in Neuroscience at Massachusetts General Hospital. He co-chairs the W3C Open Annotation Community Group, and is principal architect of the Domeo web annotation toolkit.

Stian Soiland-Reyes (http://orcid.org/0000-0001-9842-9718) is a technical software architect and researcher at the School of Computer Science, University of Manchester, UK. He is a contributor to the Open Annotation specification and the W3C Provenance Model, and one of the creators of the Wf4Ever Research Object model.

Tim Clark (http://orcid.org/0000-0003-4060-7360) is Director of Informatics at the MassGeneral Institute for Neurodegenerative Disease, Instructor in Neurology at Harvard Medical School, and Principal Investigator of the Domeo project. He is one of the founders of the W3C Open Annotation Community Group.

## *Figures*

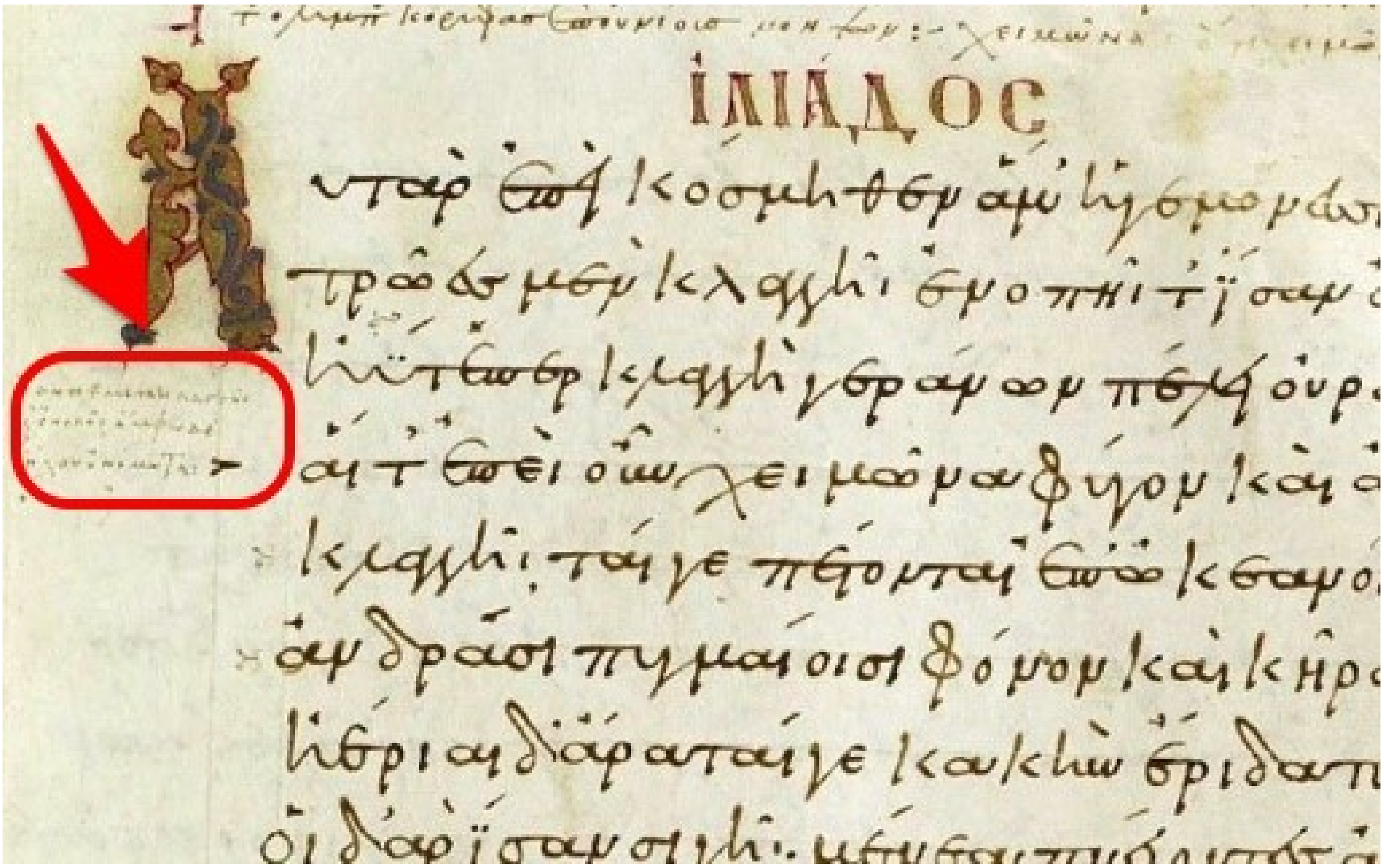

**Figure 1: Ancient annotation detail from the 500-year-old Venetus A manuscript, showing Iliad 3.1-9.** Imaged and translated, with additional digital annotations, by the Homer Multitext Project (http://www.homermultitext.org/). You can interact with Homer Multitext annotations on your iPad using this app: https://itunes.apple.com/us/app/imaging-the-iliad/id422984341?mt=8.

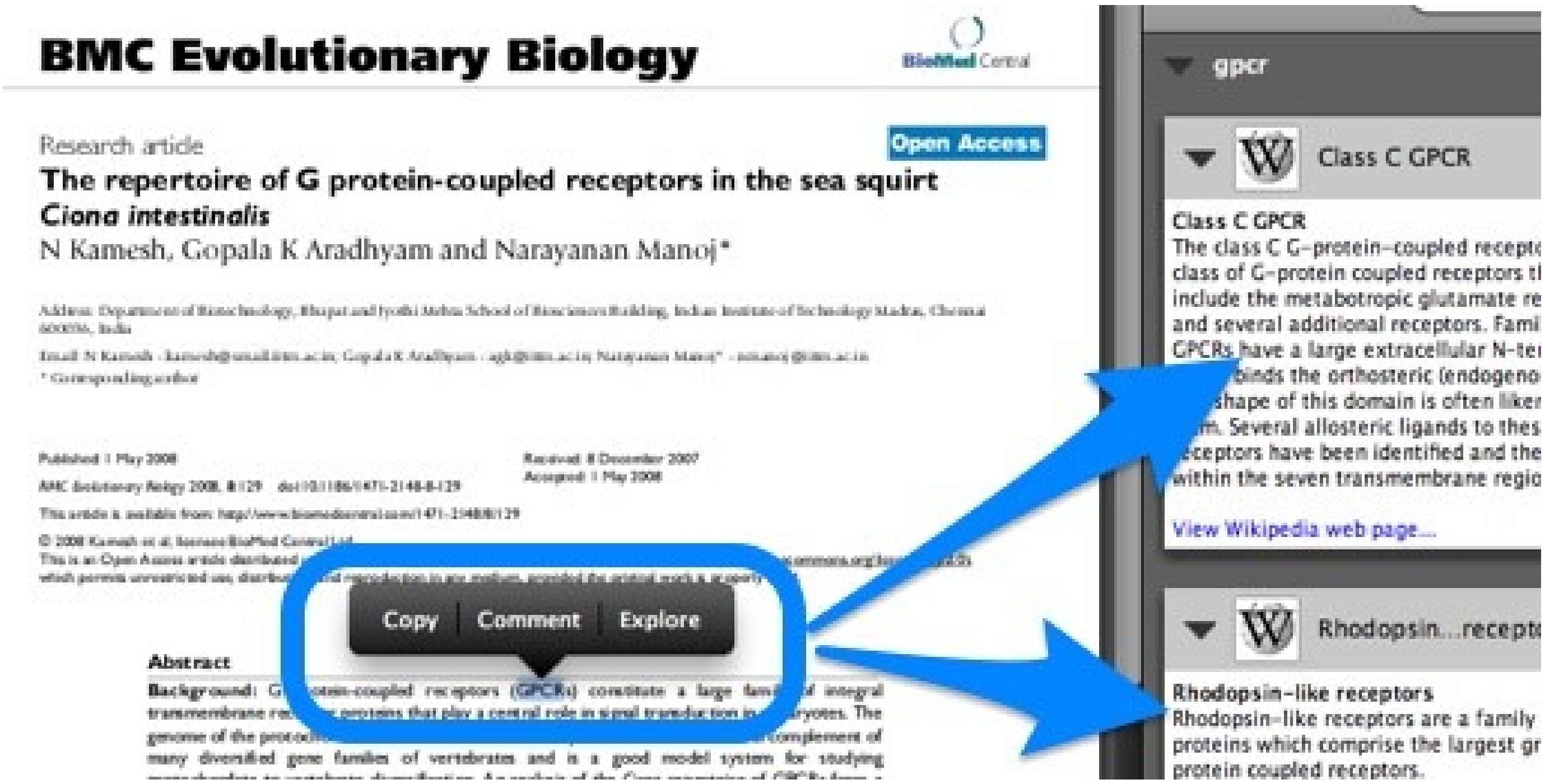

**Figure 2: Annotation of a scientific article using Wikipedia entries in the Utopia Documents system.**

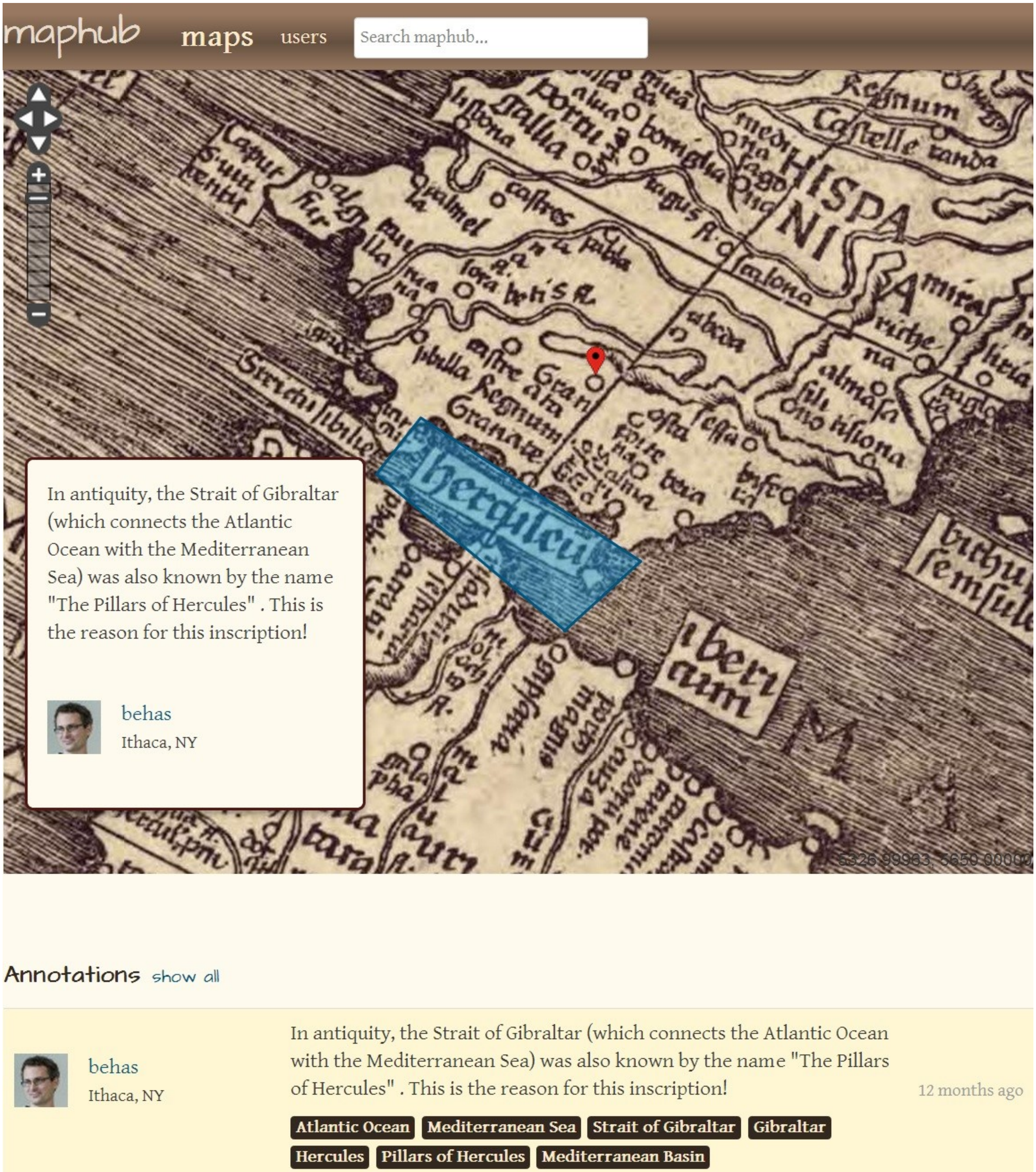

***Figure 3: Maphub* annotating a c. 1507 map image with semantic tags for the *Gibraltar* and *Hercules* entries in *DBPedia* .**

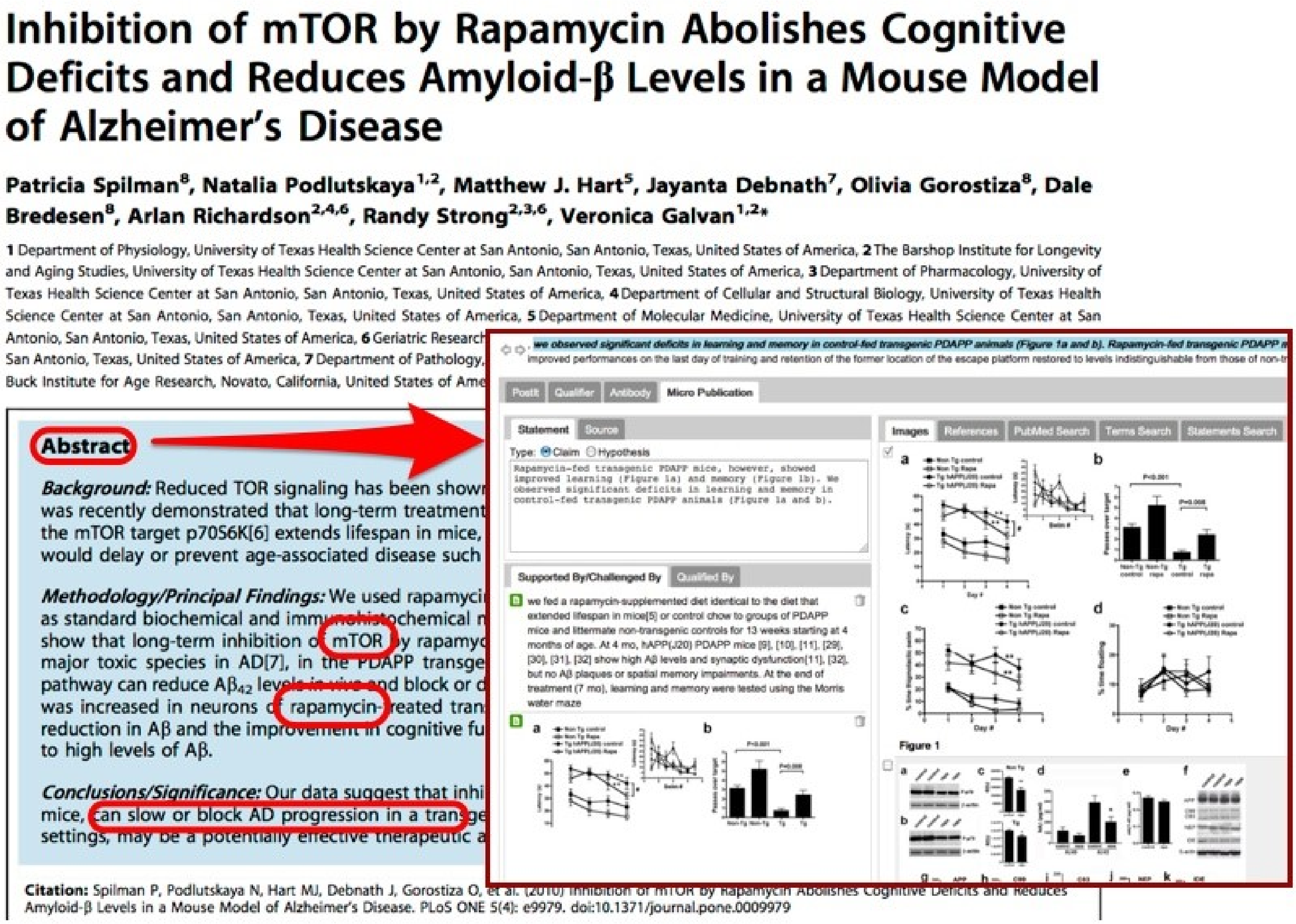

**Figure 4: Creating a *micropublication* digital summary annotation on a research article using Domeo**.

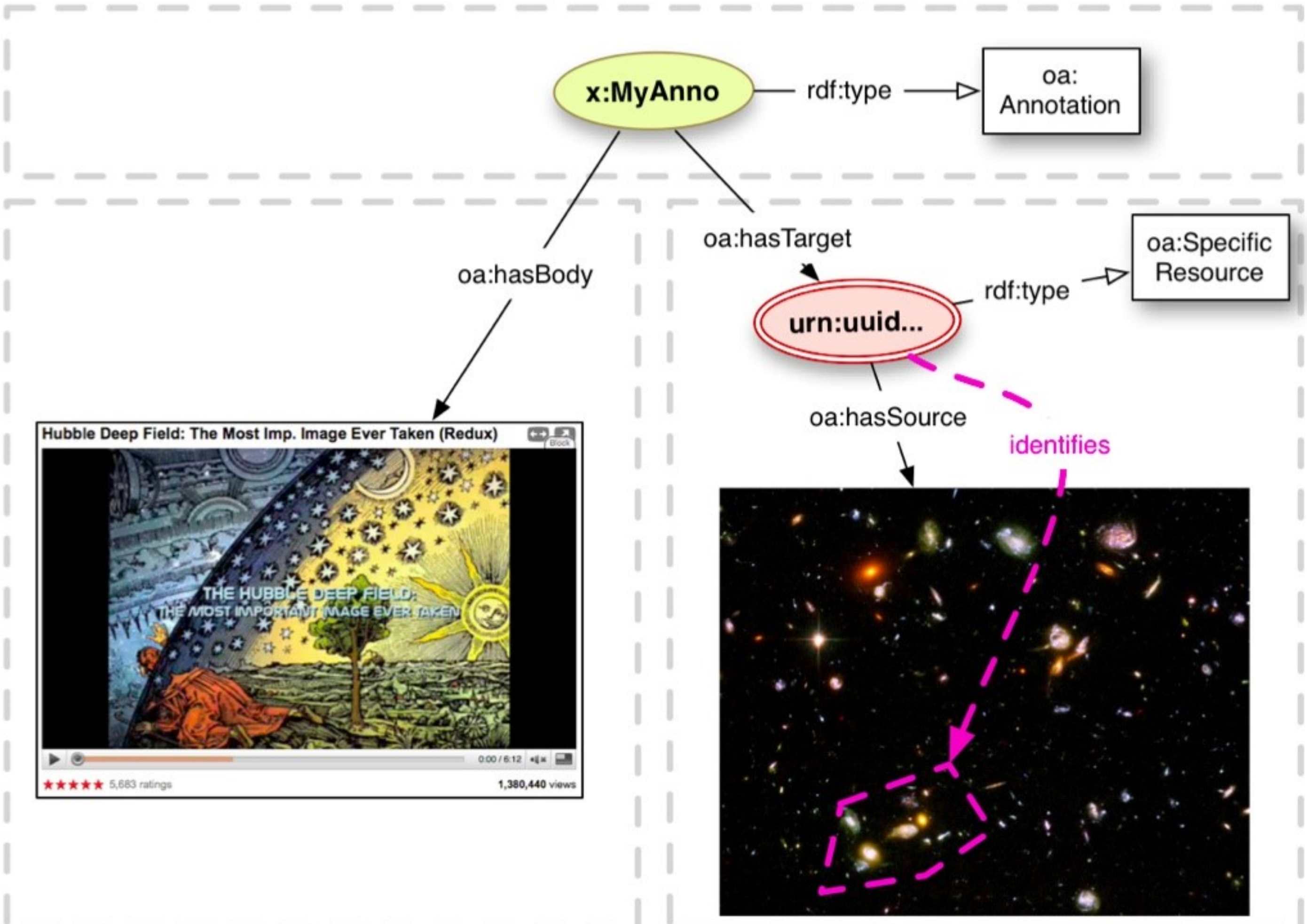

**Figure 5: Not just text: An image from the Hubble Deep Field telescope annotated using OA.** The *Target* is an image of a group of galaxies. The *Body* is a video about the image.